\documentclass[twocolumn,showpacs,preprintnumbers,amsmath,amssymb]{revtex4}

\usepackage{graphicx}
\usepackage{dcolumn}
\usepackage{bm}

\newcommand{\be}{\begin{equation}}
 \newcommand{\ee}{\end{equation}}
 \newcommand{\bea}{\begin{eqnarray}}
 \newcommand{\eea}{\end{eqnarray}}

\newcommand{\hk}{\hspace{0.1cm}}

\newcommand{\rk}{\right)}
\newcommand{\lk}{\left(}

\newcommand{\fx}{\boldsymbol{x}}
\newcommand{\fk}{\boldsymbol{k}}

\newcommand{\vk}{\vec{k}}
\newcommand{\vA}{\vec{A}}
\newcommand{\vx}{\vec{x}}

\newcommand{\vB}{\vec{B}}
\newcommand{\vE}{\vec{E}}

\renewcommand{\vec}[1]{\mbox{\boldmath$#1$\unboldmath}}

\begin{document}


\title{Dielectric function of the QCD vacuum }

\author{H. Reinhardt}
 \email{hugo.reinhardt@uni-tuebingen.de}
\affiliation{%
Institut f\"ur Theoretische Physik\\
Auf der Morgenstelle 14\\
D-72076 T\"ubingen\\
Germany
}%


\date{\today}

\begin{abstract}
It is shown that the inverse of the ghost form factor in the Hamilton approach
to Yang-Mills theory in Coulomb gauge can be
interpreted as the color dielectric function of the QCD vacuum. Furthermore the
horizon condition to the ghost form factor implies that in the infrared the QCD
vacuum is a perfect color diaelectric medium and therefore a dual
superconductor. The dielectric function is explicitly calculated within a
previously developed variational approach, using a specific ansatz for the
vacuum wave functional.
\end{abstract}

\pacs{11.10Ef, 12.38Aw, 12.38Cy, 12.38Lg}
\maketitle

\section{\label{sec1}Introduction}

Soon after the discovery of QCD it was realized that due to the confinement of
color charges the QCD vacuum must behave like a perfect (or nearly perfect)
color diaelectric medium \cite{RX1}. This picture is also realized in the MIT bag
 model \cite{RX2}
or SLAC bag model \cite{RX3}. 
However, so far, no real attempt has been undertaken to calculate the color dielectric
function of the QCD vacuum from the underlying theory. Obviously, such
calculations of vacuum properties have to be nonperturbative.

The QCD vacuum and, in particular, the confinement mechanism have been subject
to intensive studies, from which several pictures of confinement have emerged
like the dual Meissner effect \cite{RM},
 center vortex condensation \cite{RX1a}, \cite{RYZ} or
the Gribov-Zwanziger confinement mechanism \cite{R10},
\cite{R11}. Although there is evidence from lattice calculations that the various confinement pictures---in
particular the first two---are related no direct 
connection between these pictures have been established so far. In the present
paper I show that the Gribov-Zwanziger confinement scenario in Coulomb gauge
implies the dual superconductor. This connection is based on the 
observation made in sect.  III that the inverse of the ghost from factor in
Coulomb gauge represents the dielectric function of the Yang-Mills vacuum. 

Recently, progress has been made in determining the Yang-Mills vacuum wave
functional by a variational solution of the Yang-Mills Schr\"odinger equation inCoulomb gauge \cite{R4}, \cite{R5}. 
The infrared properties calculated with the wave functional
obtained in refs.\ \cite{R5}, \cite{R6} and \cite{Feuchter:2007mq} show clear
signals of confinement: an infrared divergent gluon energy, a linearly rising
static quark potential and a perimeter law for the 't Hooft loop \cite{R7}. 
Therefore, we may expect that the variationally determined vacuum wave
functional contains the essential infrared physics of the Yang-Mills vacuum. In
the present paper I shall use this wave functional to explicitly 
calculate the dielectric
function of the QCD vacuum. I will first identify the dielectric function 
 in the Hamilton approach in Coulomb gauge and discuss some of
its general properties, which do not rely on the variational approach. 
After that I will present the numerical results for
this quantity obtained by using the wave functional determined in ref. 
\cite{R6}.

\section{The Hamilton approach to Yang-Mills theory in Coulomb gauge \label{sec2}}
Consider the Hamilton formulation of Yang-Mills theory in Weyl gauge $A^a_0 (\fx)
= 0$. The electric field $E^a_i$ represents the momentum conjugate to the
spatial components of the gauge field $A^a_i$. In the canonical quantization the
electric field is replaced by the momentum operator $\Pi^a_i (\fx) = \frac{1}{i}
\frac{\delta}{\delta A^a_i (\fx)}$ and the Yang-Mills Hamiltonian reads
\be
\label{1}
H = \frac{1}{2} \int d^3 x \lk \vec{\Pi}^2 (\fx) + \vB^2 (\fx) \rk \hk ,
\ee
where $B^a_i$ is a non-abelian magnetic field.

In Weyl gauge, Gauss' law is lost from the equation of motion and has to be
imposed as a constraint to the wave functional
\be
\label{2}
\hat{D}^{a b}_i \Pi^b_i (\fx) | \psi \rangle = g \rho^a (\fx) | \psi \rangle
\hk .
\ee
Here $\hat{D}^{a b}_i = \delta^{a b} \partial_i + g\hat{A}^{a b}_i \hk , \hk \hat{A}^{a
b} = f^{a c b} A^c $ is the covariant derivative in the adjoint representation
of the gauge group ($f^{a b c}$ being the structure constant of the
gauge group and $g$ being the coupling constant)
and $\rho^a (\fx)$ denotes the ``external'' color charge density of the matter
fields. The operator on the left hand side of eq.\ (\ref{2}) is the generator of
time-independent gauge transformations and thus Gauss' law
expresses the gauge invariance of the wave functional. 

Instead of working with
gauge invariant wave functionals it is more convenient to explicitly resolve
Gauss' law by fixing the gauge. For this purpose, Coulomb gauge $\vec{\partial}
\vA^a = 0$ is a particularly convenient gauge, which will be used in the
following. In Coulomb gauge only the transversal components of the gauge field
$A = A_\perp$ are left. Splitting the momentum operator in longitudinal and
transversal parts $\Pi = \Pi_{||} + \Pi_\perp \hk , \hk \Pi_\perp = \frac{1}{i}
\delta / \delta A^a_\perp$, respectively, Gauss' law can be solved for the longitudinal part
yielding 
\be
\label{3}
\vec{\Pi}_{||} | \psi \rangle = - g\vec{\partial} (- {\bf \hat D} \boldsymbol{\partial})^{- 1} (\rho +
\rho_g) | \psi \rangle \hk ,
\ee
where $\rho^a_g = - \hat{A}^{a b}_{\perp i} \Pi^b_{\perp i}$ is the color charge density
 of the
gauge bosons and $(-{\bf \hat{D}}\boldsymbol{\partial})$ is the Faddeev-Popov kernel in Coulomb
gauge.

We are interested in the electric field which in the quantum theory is defined
as the expectation value of the corresponding (momentum) operator
\be
\label{4}
\vE = \langle \vec{\Pi} \rangle = \langle \vec{\Pi}_{||} + \vec{\Pi}_\perp
\rangle \hk .
\ee
The vacuum wave functional can be assumed to be invariant with respect to the
transformation
$A_\perp \to - A_\perp$, implying that $\langle \vec{\Pi}_\perp \rangle = 0$. 
With this result we obtain from eq.\ (\ref{3}) for the electric field
\be
\label{5}
\vE = \langle \vec{\Pi}_{||} \rangle = - \vec{\partial} \lk g \langle
(- {\bf \hat{D}}\boldsymbol{\partial})^{- 1} \rangle \rho + g \langle (- {\bf \hat{D}}\boldsymbol{\partial})^{- 1} \rho_g \rangle
\rk \hk .
\ee
Here the first term represents the response of the Yang-Mills vacuum to the
presence of the external color charges $\rho^a (\fx)$. In the absence of external
color charges $\rho^a (\fx) = 0$ the Yang-Mills vacuum should not contain any
observable color electric field, i.e. $\langle \Pi \rangle_{\rho = 0} = 0$, from
which we can conclude that the second term in eq.\ (\ref{5}) vanishes in the
Yang-Mills vacuum state. The color electric field generated by external color
charges $\rho^a(\fx)$ is therefore given by
\be
\label{6}
\vE^a (\fx) = - \vec{\partial}_{\fx} \int d^3 x' G^{a b} ({\fx}, {\fx}') \rho^b ({\fx}') \hk ,
\ee
where
\be
\label{7}
G^{a b} ({\fx}, {\fx}') = \langle (-{\bf  \hat{D}}\boldsymbol{\partial})^{- 1} \rangle ({\fx}, {\fx}')
\ee
is the ghost propagator. Let us emphasize that eq.\ (\ref{6}) has exactly the
same structure as the electric field generated by an ordinary electric charge
density $\rho (\fx)$ in classical electrodynamics
\bea
\label{8}
\vE (\fx) & = & - \vec{\partial}_{\fx} \int d^3 x' \langle {\fx} | (- \Delta)^{- 1} | {\fx}'
\rangle \rho ({\fx}') \nonumber\\
& = & - \vec{\partial}_{\fx} \int d^3 x' \frac{\rho ({\fx}')}{4 \pi | {\fx} - {\fx}'|} \hk 
\eea
except that the Green function of the Laplacian is replaced by the ghost
propagator. \footnote{Note in electrodynamics the coupling constant $e$
is absorbed into the charge units.}

\section{The dielectric function}
The ghost propagator, eq.\ (\ref{7}), is a property of the Yang-Mills vacuum (and
does not depend on the external charges). By global color invariance of the
Yang-Mills vacuum this propagator has to be color-diagonal.
 Furthermore, by translational invariance of the Yang-Mills vacuum $G^{a
b} (\vx, \vx')$ depends only on $|\vx - \vx'|$. In momentum space the ghost
propagator has therefore the following form
\be
\label{9}
g\, G^{a b} (\vk) = \delta^{a b} \frac{d (\vk)}{\vk^2} \hk .
\ee
Here we have introduced the ghost form factor $d (\fk)$, which describes the
deviations of the ghost propagator from the Green function of the Laplacian,
i.e. it embodies all the deviations of the Yang-Mills vacuum from the QED case.
With (\ref{9}) we find from eq.\ (\ref{6}) for the electric field in momentum
space
\be
\label{10}
\vE^a (\fk) = - i \vk \frac{d (\fk)}{\vk^2} \rho^a (\vk) \hk ,
\ee
where $\vE^a (\vk)$ and $\rho^a (\vk)$ denote the Fourier transform of $\vE^a
(\vx)$ and $\rho^a (\vx)$, respectively. Equation (\ref{10}) should be compared with
the expression for the ordinary electric field in a medium 
\be
\label{11}
\vE (\vk) = - i \vk \frac{1}{\epsilon (\vk) \vk^2} \rho (\vk) \hk ,
\ee
where
$\epsilon (\vk)$ is the (generalized) 
dielectric function \cite{RY} and $\rho (\vk)$ the Fourier
transform of the electric charge density. Comparing eqs. (\ref{10}) and
(\ref{11}) we can 
identify the inverse of the ghost form factor as the dielectric
function of the Yang-Mills vacuum
\be
\label{12}
\epsilon (\fk) = d^{- 1} (\fk) \hk .
\ee
In QED the Faddeev-Popov kernel of Coulomb gauge is given by the Laplacian and
the ghost form factor is $d (\fk) = 1$, so that $\epsilon (\fk) = 1$ in the QED 
vacuum, as expected.

On general grounds one expects that the ghost form factor in QCD is infrared
divergent 
\be
\label{13}
d^{- 1} (\fk = {\bf 0}) = 0 \hk . 
\ee
This is the so-called horizon condition, which is required for
confinement in the Gribov-Zwanziger confinement scenario \cite{R10},
\cite{R11}. By this
condition the dielectric function of the Yang-Mills vacuum (\ref{12}) vanishes
in the infrared 
\be
\label{14}
\epsilon (\fk = {\bf 0}) = 0
\ee
and hence the Yang-Mills vacuum is a perfect color diaelectric medium.

A perfect color diaelectric medium $\epsilon = 0$ is nothing but a dual
superconductor (where dual refers to an interchange of electric and magnetic
fields and charges). Recall in an ordinary superconductor the magnetic
permeability vanishes, $\mu = 0$, and, consequently, in a dual superconductor
$\epsilon = 0$. We thus observe that the Gribov-Zwanziger confinement scenario
is consistent with the dual Meissner effect proposed as possible confinement
scenario \cite{RM},
 \cite{R12}, and supported by lattice calculations \cite{R13}. 

Let us emphasize that the ghost propagator, by definition, depends on the gauge
chosen and only the ghost form factor of Coulomb gauge is directly related to
the dielectric function by eq. (\ref{12}). 
\begin{figure}
\includegraphics[width=0.5\textwidth]{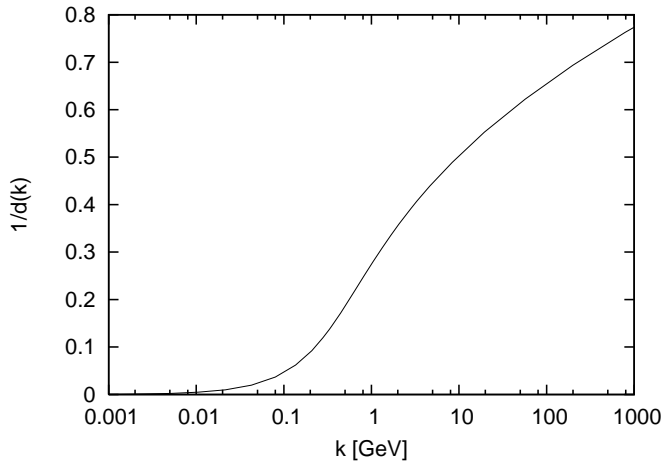}
\caption{\label{invghostnolog} The dielectric function $\epsilon (\fk)$ 
(\ref{12}) of the Yang-Mills
vacuum calculated in the variational approach \cite{R5} using the solution of
the Dyson-Schwinger equations reported in ref.\ \cite{R6}.}
\end{figure}

The ghost
form factor $d (\fk)$ has been explicitly calculated in the variational solution
of the Yang-Mills Schr\"odinger equation in Coulomb gauge \cite{R5}, \cite{R6}.
 Fig. 1
shows the result for the dielectric function (\ref{12}) obtained from the
solutions to the Dyson-Schwinger equations reported in \cite{R6}. 
It has the expected behavior: at zero momentum it vanishes by the
horizon condition, while
for $k \to \infty$ it diverges logarithmically due to the anomalous dimension of
the ghost propagator. This behavior of $\epsilon (\fk) = d^{- 1} (\fk)$ is a
manifestation of anti-screening in Yang-Mills theory. (Ordinary Debye screening,
which turns the Coulomb potential into a Yukawa potential, produces a
dielectric function
\be
\epsilon (\fk) = \frac{m^2 + \fk^2}{\fk^2} \hk ,
\ee
where $m$ is the inverse screening distance. This dielectric function is
divergent at $k \to 0$ and approaches the QED vacuum value $\epsilon = 1$ for $k
\to \infty$).

Let us emphasize that the vanishing of the dielectric function in the infrared
is a consequence of the horizon condition (\ref{13}), which is an intrinsic
feature of the Gribov-Zwanziger confinement mechanism. This condition 
has been imposed on
the solution to the Dyson-Schwinger equations following from the variational
approach \cite{R5}, \cite{R6}. While in $D = 3 + 1$, in principle, solutions to
these equations with an infrared finite ghost form factor can be found
\cite{Epp+07}, in $D = 2 + 1$ these equations allow only for solutions satisfying
the horizon condition \cite{Feuchter:2007mq}. 
The ghost form factor in Coulomb gauge has also been calculated on the lattice
in both $D = 3 + 1$, 
\cite{R8}, \cite{R9} and $D = 2 + 1$ \cite{RZ}. 
Unfortunately the lattices used so far in $D = 3 + 1$ 
are not large enough to really
penetrate the infrared regime $k < \sqrt{\sigma}$ ($\sigma-$string tension).
In the momentum regime where reliable lattice data are available, there is a
reasonable agreement between lattice data \cite{R9},  \cite{RZ} 
and the continuum results \cite{R5}, \cite{R6}, \cite{Feuchter:2007mq}. 
This refers, in particular, to
$D = 2 +1$ dimensions, where larger lattices can be used.
The $2 + 1$ dimensional lattice calculation  \cite{RZ} give strong evidence for
an infrared divergent ghost form factor and are in quite satisfactory agreement
with the continuum results \cite{Feuchter:2007mq}.

In the Zwanziger-Gribov confinement scenario the horizon condition is understood
to arise from the field configuration on or near the Gribov horizon, where the
Faddeev-Popov kernel develops a zero eigenvalue. One can show analytically
\cite{Greensite:2004ur} that in Landau as well as Coulomb gauge, center vortices
and magnetic monopoles lie on the Gribov horizon. The latter configurations are
responsible for the dual Meissner effect. Furthermore, when the center vortex
configurations are removed from the Yang-Mills lattice ensemble
\cite{Gattnar:2004bf}, \cite{Greensite:2004ur} the infrared singular behavior
of the ghost from factor is lost. In this sense the Gribov-Zwanziger confinement
scenario does not only imply the dual Meissner effect but is also linked to the
center vortex condensation picture of confinement.
\section{Conclusions}
I have shown that the ghost form factor in the Hamilton approach to Yang-Mills
theory in Coulomb gauge can be interpreted as the inverse of the dielectric
function of the Yang-Mills vacuum. Consequently the horizon condition, a
necessary requirement for the Gribov-Zwanziger confinement scenario, implies that
in the infrared the Yang-Mills vacuum behaves like a perfect color diaelectric
medium, which, in fact, represents a dual superconductor. In this way I have
shown that the Gribov-Zwanziger confinement scenario implies the dual
Meissner effect.

\begin{acknowledgments}
The author is grateful to W. Schleifenbaum for providing the plot of fig.\ 1.
He also thanks G. Burgio, M. Quandt, W. Schleifenbaum and P. Watson for a
critical reading of the manuscript and useful comments.
 This work was supported in part by DFG under DFG-Re856/6-1
and DFG-Re856/6-2.
\end{acknowledgments}

\end{document}